\documentclass[conference]{IEEEtran}
\IEEEoverridecommandlockouts
\usepackage{cite}
\usepackage{amsmath,amssymb,amsfonts}
\usepackage{algorithmic}
\usepackage{graphicx}
\usepackage{textcomp}

\usepackage{float}
\usepackage{threeparttable} 
\usepackage{soul}
\usepackage{hyperref}
\usepackage[table,xcdraw]{xcolor}

\usepackage{multirow}
\usepackage{threeparttable}
\usepackage{breqn}
\usepackage{tikz}

\def\BibTeX{{\rm B\kern-.05em{\sc i\kern-.025em b}\kern-.08em
    T\kern-.1667em\lower.7ex\hbox{E}\kern-.125emX}}

\newcommand*\filledcircled[1]{\tikz[baseline=(char.base)]{
    \node[shape=circle, draw, inner sep=0.25pt, fill=black, text=white] (char) {#1};}}

\begin{document} 
\title{A Comparative Analysis of Microrings Based Incoherent Photonic GEMM Accelerators}

\author{\IEEEauthorblockN{Sairam Sri Vatsavai, Venkata Sai Praneeth Karempudi, Oluwaseun Adewunmi Alo, and Ishan Thakkar }
\IEEEauthorblockA{\textit{Department of Electrical and Computer Engineering, University of Kentucky, Lexington, USA } \\
ssr226@uky.edu, kvspraneeth@uky.edu, seun.alo@uky.edu, and igthakkar@uky.edu}}
% \and
% \IEEEauthorblockN{Venkata Sai Praneeth Karempudi }
% \IEEEauthorblockA{\textit{Electrical and Computer Engineering} \\
% \textit{University of Kentucky}\\
% Lexington, USA  \\
% kvspraneeth@uky.edu}
% \and
% \IEEEauthorblockN{Alo Oluwaseun }
% \IEEEauthorblockA{\textit{Electrical and Computer Engineering} \\
% \textit{University of Kentucky}\\
% Lexington, USA  \\
% seun.alo@uky.edu}
% \and
% \IEEEauthorblockN{Ishan Thakkar }
% \IEEEauthorblockA{\textit{Electrical and Computer Engineering} \\
% \textit{University of Kentucky}\\
% Lexington, USA  \\
% igthakkar@uky.edu}
% }

\maketitle

\begin{abstract}
Several analog photonic architectures based on microring resonators (MRRs) have been proposed to accelerate general matrix-matrix multiplications (GEMMs) that compose deep neural networks with exceptional throughput and energy efficiency. To implement GEMM functions, these MRR-based architectures, in general, manipulate optical signals in five different ways: (i) Splitting (copying) of multiple optical signals to achieve a certain fan-out, (ii) Aggregation (multiplexing) of multiple optical signals to achieve a certain fan-in, (iii) Modulation of optical signals to imprint input values onto analog signal amplitude, (iv) Weighting of modulated optical signals to achieve analog input-weight multiplication, (v) Summation of optical signals. Prior accelerators undertake the first four ways of signal manipulation in an arbitrary order ignoring the possible impact of the order of these manipulations on their performance. In this paper, we conduct a detailed analysis of accelerator organizations with three different orders of these manipulations: (1) Modulation-Aggregation-Splitting-Weighting (MASW), (2) Aggregation-Splitting-Modulation-Weighting (ASMW), and (3) Splitting-Modulation-Weighting-Aggregation (SMWA). We show that these organizations affect the crosstalk noise and optical signal losses in different magnitudes, which renders these organizations with different levels of processing parallelism at the circuit level, and different magnitudes of throughput and energy-area efficiency at the system level. Our evaluation results for four CNN models show that the SMWA organization achieves up to 4.4$\times$, 5$\times$, and 5.2$\times$ better throughput, energy efficiency, and area efficiency, respectively, compared to the ASMW and MASW organizations on average.
\end{abstract}

\section{Introduction}
Deep Neural Networks (DNNs) achieve high inference accuracy, which has revolutionized their use in various artificial intelligence tasks, such as image recognition, language translation, and autonomous driving \cite{dnnapplications1,dnnapplications2}. However, DNNs are computationally intensive because they are typically composed of inherently abundant linear functions such as general matrix-matrix multiplication (GEMM). 
%Convolutional Neural Networks (CNNs) are specific types of DNNs \cite{cnnapplication}. CNNs are computationally intensive, and hence, require a long inference time. In CNNs, around 80\% of the total processing time is taken by convolution operations.
The need to tackle the rapidly increasing computing demands of the abundant GEMM functions of DNNs has pushed for highly customized hardware GEMM accelerators \cite{Baischer2021,sze2017efficient}. 
%For hardware acceleration of convolution operations, they can be converted into general matrix-matrix multiplication (GEMM) operations \cite{sze2017efficient}.
%The order of mapping determines the dataflow and the performance of the hardware varies with dataflow. 
%Often, for efficient and swift hardware-based acceleration, CNNs are quantized to have integer input/weight parameters \cite{krishnamoorthi2018quantizing}. 
Among GEMM accelerators in the literature, silicon-photonic GEMM accelerators have shown great promise to provide unparalleled parallelism, ultra-low latency, and high energy efficiency \cite{holylight,squeezelight,deapcnn,karen2020proceeding,crosslight}. A silicon-photonic GEMM accelerator typically consists of multiple Dot Product Units (DPUs) that perform a total of \textit{M} dot product operations in parallel of \textit{N}-size each. Several DPU-based optical GEMM accelerators have been proposed in prior works. Among them, the Microring Resonator (MRR)-enabled analog DPU-based accelerators (e.g., \cite{holylight,deapcnn,squeezelight,pixel,crosslight,tait2017}) have shown disruptive performance and energy efficiencies, due to the MRRs' compact footprint, low dynamic power consumption, and compatibility with cascaded dense-wavelength-division multiplexing (DWDM).

A typical DPU employs five blocks of optical components to manipulate optical signals in five different ways. (1) A splitting block for splitting (copying) \textit{N} optical signals in \textit{M} ways to achieve a fan-out of \textit{M} per optical signal; (2) An aggregation block for aggregation (multiplexing) of \textit{N} optical signals per waveguide to achieve a fan-in of \textit{N} per waveguide; (3) A  modulation block for modulation of optical signals to imprint input values onto them; (4) A weighting block for weighting of modulated optical signals to achieve analog input-weight multiplication; (5) A summation block to perform summation of optical signals. Prior accelerators arrange these optical signal manipulation blocks within a DPU in an arbitrary order, resulting in different DPU organizations. Different DPU organizations incur different severity levels of various optical crosstalk effects and signal losses, causing different amounts of optical power penalty across different DPU organizations. This variation in optical power penalty causes different DPU organizations to achieve different values of \textit{N} (fan-in degree/DPE size) and \textit{M} (fan-out degree/count of parallel DPEs). This is because the achievable peak values of \textit{N} and \textit{M} highly depend on the available optical power budget in the DPU, which in turn is determined by the incurred power penalty in the DPU \cite{cases2022}. It can be intuitively hypothesized that different values of \textit{N} and \textit{M} would render different DPU organizations with different levels of processing parallelism at the circuit level, and different magnitudes of throughput and energy-area efficiency at the system level. However, no prior work has tested this hypothesis. We address this shortcoming in this paper.  

%the optical power budget per DPU to be whittled down by different amounts. The achievable dot product size (\textit{N}), bit precision (B), and supported parallel dot product operations (M) in an optical DPU are contingent on the available optical power budget. This budget is constrained by the balanced photodetector's lower-bound sensitivity and the upper-bound injected laser power. Within an optical DPU, it is necessary to account for all power penalties, including those resulting from various losses like crosstalk noise and optical signal losses. The remaining power is allocated to support DPU size N(=M) and B. However, prior works have not explored the impact of DPU organization on overall losses, which, in turn, affects scalability and bit precision achieved by a DPU at the circuit level. The existing literature also lacks insight into the affects of organization on system-level performance, including throughput and energy-area efficiency.

%To address these shortcomings, we categorize previous analog optical MRR-based DPUs into three groups based on the organization order of optical signal manipulation blocks. We then conduct a comprehensive circuit-level analysis to assess the impact of the organization on overall losses within a DPU. Leveraging the insights from our analysis, we evaluate the scalability limits of each organization at the circuit level and assess system-level performance, including throughput and energy-area efficiency.          

Our key contributions in this paper are summarized below.
\begin{itemize}
\item We classify the DPU organizations from prior work into three categories, namely ASMW, MASW, and SMWA;
\item We analyze and discuss the impact of different DPU organizations on various optical crosstalk effects and signal losses;
\item We perform a comparative analysis of the impact of different DPU organizations on the scalability of achievable \textit{N} and \textit{M} values at different bit precision values;
%\item  We perform the scalability analysis of DPUs of ASMW, MASW, and SMWA categories to find the maximum achievable DPU size across a range of bit precision levels for different operating data rates (Section \ref{sec4})
%\item We evaluate the impact of MRR-based DPU organization on GEMM execution performance and energy consumption (Section \ref{sec4})
\item We implement and evaluate ASWM, MASW, and SMWA organizations at the system-
level with our in-house simulator, and report the  performance in terms of throughput (FPS), energy-efficiency (FPS/W), and area-efficiency (FPS/W/mm$^2$), for the inferences of
four CNNs.
\end{itemize}

\section{Preliminaries}

\subsection{Processing of CNNs on Hardware Accelerators}
\begin{figure}[h]
  \centering
  \includegraphics[scale=0.34]{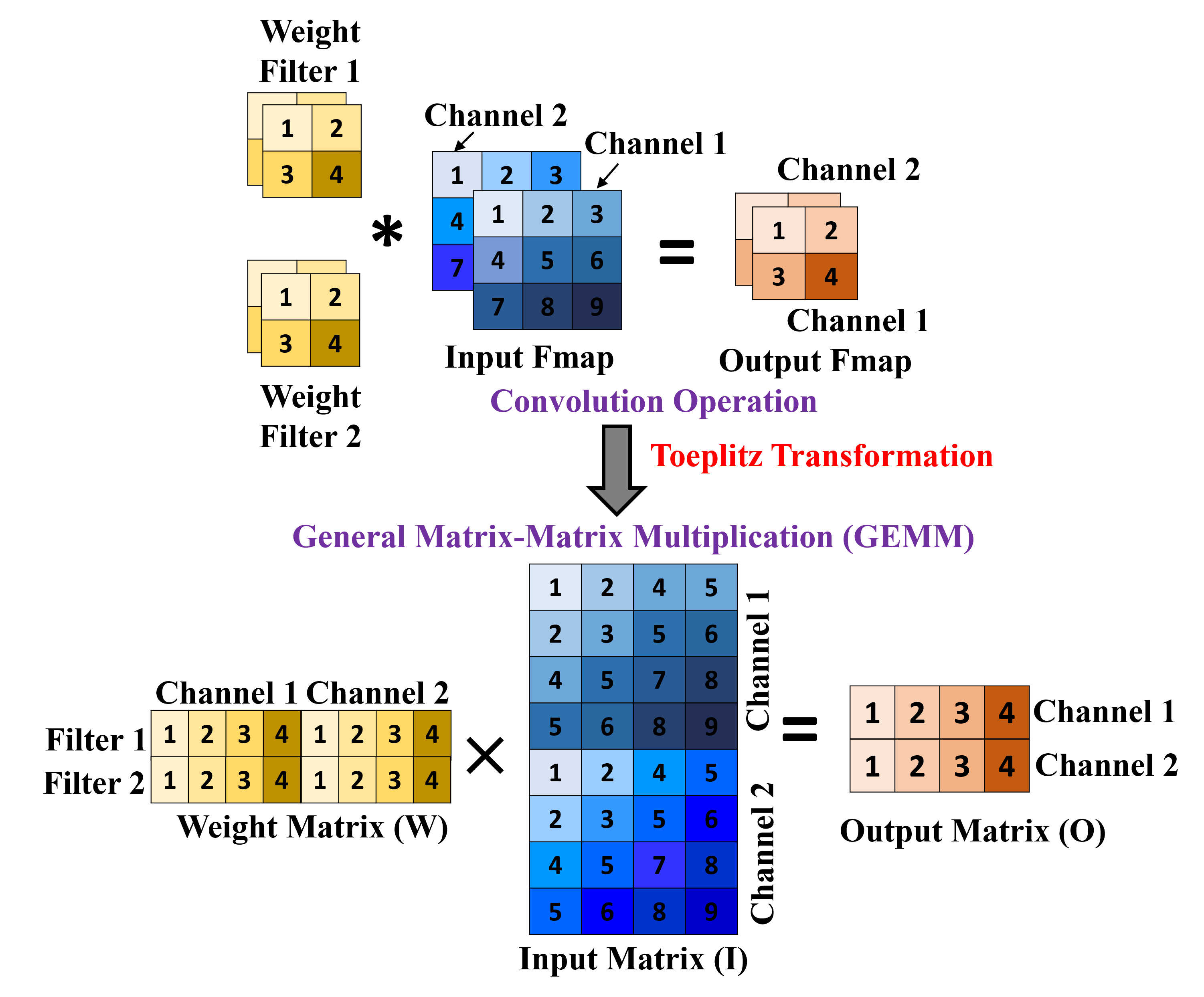}
  \caption{Convolution operation at a convolution layer with two weight filters and one input feature map (Fmap) having two channels is transformed into a GEMM operation between input matrix \textit{\textbf{I}} and weight matrix \textit{\textbf{W}}.} 
  \label{fig1}
\end{figure}

In CNNs, the major computational requirement arises from convolutional layers. These layers involve convolution operations that can be converted to General Matrix-Matrix Multiplication (GEMM) operations using the Toeplitz matrix or the im2col transformation \cite{sze2017efficient,pytorchUnfold}. As shown in Fig. \ref{fig1}, the input feature map (Fmap) belonging to a convolution layer is unfolded into the matrix \textbf{\textit{I}}. The weight filters of the convolution layer are flattened and stacked to form the weight matrix (\textbf{\textit{W}}). The GEMM operation between \textbf{\textit{I}} and \textbf{\textit{W}} gives the resultant output matrix (\textbf{\textit{O}}). On conventional CPUs/GPUs, GEMM operations are mapped and executed using basic linear algebra subprograms (BLAS) or Cuda BLAS (cuBLAS) \cite{BLAS,CUBLAS}. However, conventional CPUs/GPUs cannot efficiently meet the exponentially growing computational demand of modern CNNs. To meet this demand, both industry and academia have proposed various dedicated GEMM accelerators \cite{tpu, crosslight, deapcnn, albireo}, tailored to process CNNs with better performance and energy efficiency.

\subsection{Related Work on Optical GEMM Accelerators}
% Electronic ASICs have traditionally been the preferred choice for implementing CNN accelerators \cite{tpu}. In light of the deceleration of Moore's Law, coupled with the exponential growth in CNN complexity, electronic ASIC accelerators are struggling to meet the processing speed and energy efficiency demands for large-scale deployment of complex CNN models. To tackle this challenge, both industry and academic researchers are now investigating innovative more-than-Moore technologies that can offer consistently faster and energy-efficient hardware solutions for CNN acceleration in the foreseeable future. Among various technologies, silicon photonics stands out as a promising candidate, offering unparalleled parallelism, ultra-low latency, and high energy efficiency \cite{cansu2021,albireo,crosslight}. Accelerators based on silicon photonics harness linear photonic phenomena, such as optical transmission and optical signal superposition within photonic integrated circuits \cite{cansu2021,deapcnn}, to accelerate CNN inference. This acceleration results in remarkably fast processing speeds and sub-nanosecond input-to-output latency, following an O(1) scaling law. 

To accelerate CNN inferences with low latency and low energy consumption, prior work has demonstrated various GEMM accelerators based on photonic integrated circuits (PICs) (e.g., \cite{holylight,crosslight,deapcnn,adept,sconna}). Typically, PIC-based GEMM accelerators consist of multiple dot product units (DPUs), and each DPU can perform a GEMM operation on multiple constituent dop product elements (DPEs) as parallel dot product operations among the rows of the matrix \textit{\textbf{I}} and the columns of the matrix \textit{\textbf{W}}. Some accelerators implement digital DPUs (e.g., \cite{pixel,albireo,sconna}), whereas some others employ analog DPUs (e.g., \cite{holylight,crosslight,deapcnn, cases2022}). Different DPU implementations employ MRRs (e.g., \cite{holylight,deapcnn,crosslight,tait2020photonic,sconna}) or MZIs (e.g., \cite{adept,mzicomplex2021}) or both (e.g., \cite{pixel}, \cite{albireo}). 
%Analog DPUs can be further classified as incoherent DPUs (e.g., \cite{holylight,crosslight,deapcnn}) or coherent DPUs (e.g., \cite{coherentrayhamerly2019,coherentZhao2019,coherentnature21}). To set and update the values of the individual input and weights used for dot product operations, the incoherent DPUs utilize the analog power amplitudes of optical signals, whereas the coherent DPUs utilize the electrical field amplitude and phase. The coherent DPUs achieve low inference latency, but they suffer from control complexity, high area overhead, low scalability, low flexibility, high encoding noise, and phase error accumulation issues \cite{limitsCohorent}. 
Among these, the accelerators based on MRRs-enabled incoherent DPUs achieve better scalability and lower footprint, because they use PICs that are based on compact MRRs \cite{deapcnn}, unlike the coherent DPUs that use PICs based on bulky MZIs. Various state-of-the-art PIC-based optical GEMM accelerators are well discussed in survey papers \cite{acceleratorssurvey,acceleratorsurvey2Sudeep,acceleratorsurvey3Bhavin}.
%Because of the inherent advantages of MRR-enabled incoherent DPUs, there is an impetus to design more energy-efficient and scalable CNN accelerators employing MRR-enabled incoherent (analog) DPUs.

\section{Organizations of MRR-based GEMM Accelerators}\label{sec3}
S. S. Vatsavai et al. in \cite{cases2022} categorized the organizations of optical MRR-based analog DPUs from prior works into two groups: Aggregate-Modulate-Modulate (AMM) and Modulate-Aggregate-Modulate (MAM) \cite{cases2022}. Here, the term "aggregate" refers to the aggregation of multiple wavelength channels into a single photonic waveguide through wavelength division-multiplexing (WDM). The first 'Modulate' refers to the modulation of optical wavelength channels with input values, and the second 'Modulate' refers to the modulation (weighting) of input-modulated optical wavelength signals with weight values. This categorization classifies prior MRR-based DPUs into AMM \cite{deapcnn,robin,crosslight} and MAM \cite{holylight,yang2013,lukasscalability} classes. However, this categorization does not consider the spectrally hitless DPU organization proposed in \cite{karen2020proceeding}. In this paper, we bridge this gap and also improve the comprehensibility of classification categories based on the order of various optical channel manipulation blocks present in MRR-based GEMM accelerators. The details of these manipulation blocks and their different organizations are discussed next. 

\begin{figure*}[] 
    \centering
    \includegraphics[width=\linewidth]{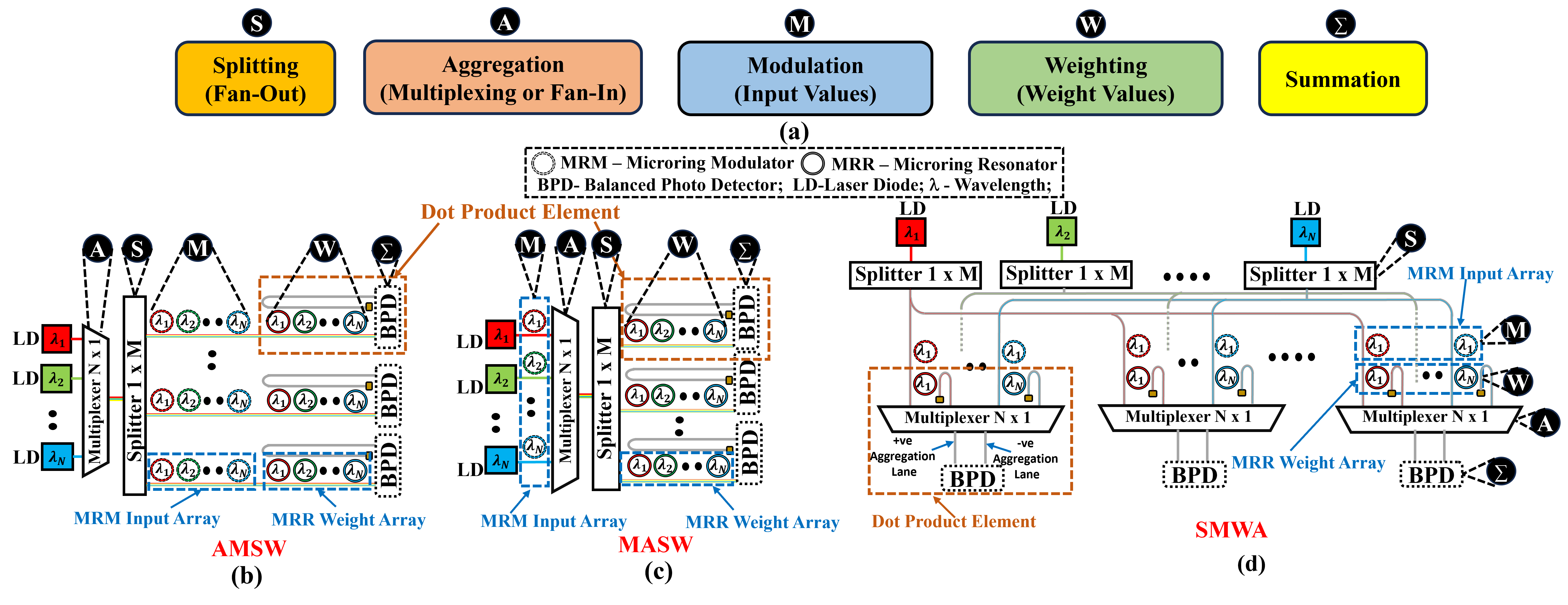}
    \caption{(a) Common optical signal manipulation blocks found in optical DPUs. Illustration of common incoherent photonic DPU organizations;   (b) AMSW DPU, (c) MASW DPU, and (d) SMWA DPU.}
    \label{fig2}
\end{figure*}

\subsection{Description of Various Blocks that Maniplate Optical Channels in MRR-based GEMM Accelerators}
%Prior MRR-based GEMM accelerators schedule the GEMM operation between input matrix \textit{\textbf{I}} and weight matrix \textit{\textbf{W}} onto their optical DPUs as dot product operations. Each DPU consists of multiple dot product elements (DPEs), and each DPE performs a dot product operation.
Every DPU in an optical GEMM accelerator employs a total of \textit{N} laser diodes that generate \textit{N} optical wavelength channels ($\lambda_1-\lambda_N$). These optical wavelength channels are manipulated by five different blocks in the DPU, as shown in Fig. \ref{fig2}(a). \textit{(i)} The splitting block (\filledcircled{S}) splits the optical power of the \textit{N} optical wavelength channels in \textit{M} copies, with each copy supporting one DPE. Thus, a total of \textit{M} DPEs are supported. \textit{(ii)} Aggregation block (\filledcircled{A}) multiplexes \textit{N} wavelength channels into a single waveguide within a DPE to achieve fan-in of \textit{N} to send \textit{N} wavelength channels concurrently toward a balanced photodetector (BPD). \textit{(iii)} Modulation block \filledcircled{M} modulates \textit{N} or \textit{N}$\times$\textit{M} wavelength channels to imprint a sequence of input values on each of the channels. These input-sequence-imprinted wavelength channels are referred to as optical signals. Each optical signal will thus be a temporal train of optical symbols, with each symbol representing a value from the input sequence. \textit{(iv)} Weighting block (\filledcircled{W}) applies weighting to the input-imprinted wavelength channels (i.e., optical signals). After weighting, each symbol carried in an optical signal represents the product of an input value and a weight value. Therefore, each such weighted optical signal is referred to as an optical product signal, with each temporal symbol of the signal representing a product value. \textit{(v)} Summation block (\filledcircled{$\Sigma$}) employs incoherent superposition at each BPD to perform symbol-wise summation of the \textit{N} optical product signals that are sent to the BPD. The BPD generates a resultant photocurrent signal, each temporal symbol of which provides a dot product of size \textit{N} (i.e., summation of \textit{N} product symbols/values). This photocurrent signal is fed to a transimpedance amplifier, followed by an analog-to-digital converter, to obtain the sequence of \textit{N}-sized dot-product results in the digital format \cite{cases2022}. The \filledcircled{$\Sigma$} block involves a total of \textit{M} BPDs corresponding to \textit{M} DPEs, thereby generating \textit{M} parallel dot product results every symbol cycle. In a DPU, generally, the five blocks in the DPU are arranged in a way so that the \filledcircled{$\Sigma$} block comes towards the end of the input-to-output propagation path of optical channels/signals within the DPU, and the \filledcircled{M} block always comes before the \filledcircled{W} block. However, different DPU organizations can differ in the way the permutational order of the blocks \filledcircled{M}, \filledcircled{W}, \filledcircled{S}, and \filledcircled{A} appear in them. Based on this order of these blocks, we classify different DPU organizations found in prior works on MRR-based GEMM accelerators into three categories: ASMW, MASW, and SMWA. Each of these organizations is explained below.

\subsubsection{ASMW DPU Organization} \label{sec3d}
Fig. \ref{fig2}(b) illustrates a DPU of ASMW organization. First, the \filledcircled{A} block appears, wherein \textit{N} wavelength channels ($\lambda_1-\lambda_N$) from LDs are multiplexed into a single waveguide. Then, in the \filledcircled{S} block, the power of each wavelength channel is equally split into \textit{M} waveguides to generate a total of \textit{N}$\times$\textit{M} parallel wavelength channels. These wavelength channels then encounter the \filledcircled{M} and \filledcircled{W} blocks in that order. The \filledcircled{M} block (\filledcircled{W} block) employs one array of \textit{N} input MRMs (\textit{N} weight MRRs) per waveguide. At the output of the \filledcircled{W} block, a total of \textit{N}$\times$\textit{M} optical product signals emerge (\textit{N} signals per waveguide in \textit{M} parallel waveguides), which are sent to the \filledcircled{$\Sigma$} block containing \textit{M} parallel BPDs.

%to the sa\textit{M} waveguides propagate \textit{M} copies (one copy pe waveguide) of \textit{N} wavelength channels towards \textit{M} DPEs in parallel.  to ultimately make them reach the \filledcircled{$\Sigma$} block. Along their propagation path, the N wavelength channels encounter \filledcircled{M} block the   to generate \textit{M} copies. Each waveguide with \textit{N} wavelength channels is modulated by one array of \textit{N} MRM input array at \filledcircled{M} block. This \textit{N} input modulated optical channels are weighted by \textit{N} MRR weighting array to generate \textit{N} optical product signals, which are sent to BPD in  to generate dot product result. The \textit{M} BPDs of \filledcircled{$\Sigma$} block produces \textit{M} dot product results.

\subsubsection{MASW DPU Organization}
Fig. \ref{fig2}(c) illustrates a DPU of MASW organization. Here, the \textit{N} optical wavelength channels generated by LDs are coupled into \textit{N} parallel waveguides (one channel per waveguide). Then, the \filledcircled{M} block appears wherein each of the \textit{N} waveguides couples with one input MRM. These \textit{N} input MRMs in the \filledcircled{M} block generate \textit{N} optical signals, which are then aggregated into a single waveguide in the subsequently appearing \filledcircled{A} block. Then, in the \filledcircled{S} block, these \textit{N} optical signals are split into \textit{M} waveguides to generate their \textit{M} copies, with each copy feeding a DPE. These N$\times$M parallel optical signals then encounter the \filledcircled{W} block, which employs an array of \textit{N} weighting MRRs per waveguide. Consequently, at the output of the \filledcircled{W} block, a total of \textit{N}$\times$\textit{M} optical product signals emerge, which are sent to the \filledcircled{$\Sigma$} block containing \textit{M} parallel BPDs.

%Thus, a total of N \textbf{} modulates  modulated with input values at \filledcircled{M} block with one MRM input array of \textit{N} MRMs. The \filledcircled{A} block multiplexes the \textit{N} input modulated wavelength channels into a single waveguide. Next, the \filledcircled{S} block equally splits the power of \textit{N} input modulated channels into \textit{M} waveguides. At each waveguide, a \textit{N} MRR weighting array of \filledcircled{W} block generates \textit{N} optical product signals. Finally, the BPD in \filledcircled{$\Sigma$} block accumulates these \textit{N} optical product signals to generate the dot product result.   

\subsubsection{SMWA DPU Organization}
Fig. \ref{fig2}(d) illustrates a DPU of SMWA organization. In this organization, the \filledcircled{S} block appears first which splits the optical power of each of the \textit{N} wavelength channels from LDs equally into \textit{M} separate waveguides by using a total of \textit{N} $1\times M$ splitters. Thus, a total of \textit{N}$\times$\textit{M} wavelength channels and \textit{N}$\times$\textit{M} waveguides emerge, with each waveguide propagating only a single wavelength channel. Then, the \filledcircled{M} and \filledcircled{W} blocks appear in that order, with both blocks containing a single MRM-MRR pair coupled to each of the \textit{N}$\times$\textit{M} waveguides. At the output of the \filledcircled{W} block, a total of \textit{N}$\times$\textit{M} optical product signals emerge (each signal in a dedicated waveguide), which are sent to the \filledcircled{A} block. There, a total of \textit{M} \textit{N}$\times$\textit{1} multiplexers are used to aggregate the optical product signals in a total of \textit{M} pairs of aggregation lanes (each pair propagating \textit{N} optical product signals). These \textit{M} pairs of aggregation lanes feed the \filledcircled{$\Sigma$} block containing \textit{M} parallel BPDs.Table \ref{classification} reports the classification of prior MRR/MZI-based GEMM accelerators based on their DPU organization. 

\begin{table}[]
\centering

\caption{ \textcolor{red}{Classification of prior MRR/MZI-based analog accelerators based on their DPU organizations.}}
\label{classification}
\resizebox{\linewidth}{!}{%
\begin{tabular}{|c|c|}
\hline
   \textbf{DPU Organization}           & \textbf{MRR-based DPU Architectures} \\ \hline
\textbf{ASMW} &       Crosslight\cite{crosslight}, DEAPCNN \cite{deapcnn}, Robin \cite{robin},  RAMM \cite{cases2022}               \\ \hline
\textbf{MASW} &       Holylight \cite{holylight}, Yang \cite{yang2013}, Al-Qadasi\cite{lukasscalability}, PCNNA \cite{pcnna}, RMAM \cite{cases2022}                \\ \hline
\textbf{SMWA} &           Hitless \cite{karen2020proceeding}, ADEPT \cite{adept}, Albireo\cite{alberio}           \\ \hline
\end{tabular}}
\end{table}
%for each set of \textit{N} waveguides carrying \textit{N} distinct wavelength channels ($\lambda_1-\lambda_N$), an MRM input array spanning across \textit{N} waveguides imprints input values onto wavelength channels. Subsequently, at \filledcircled{W} block, an \textit{N} MRR weight array weights the input imprinted wavelength channels to generate optical product signals. This organization of the MRM input array and MRR weight array is referred to as spectrally hitless organization \cite{karen2020proceeding}. Furthermore, within the MRM weight array, based on the polarity, the optical product signal is generated at MRR through port or drop port. Next, at the \filledcircled{A} block, the optical product signals from the MRR weight array are sent to \textit{N$\times$1} multiplexer. The multiplexer aggregates positive and negative optical product signals into positive and negative aggregation lanes, respectively. The \filledcircled{$\Sigma$} block generates the dot product result.  

\subsection{Motivation}
The performance achieved by the MRR-based DPUs is largely dependent on three parameters (1) The maximum achievable value of \textit{N} (fan-in degree/DPE size). Often, the achievable value of \textit{N} for photonic DPUs is less than the dot product size requirement of GEMM operations corresponding to CNN models \cite{cases2022}. In that case, a DPU breaks the dot product into smaller DPU-compatible chunks and generates intermediate results known as partial sums (\textit{psums}). These \textit{psums} are later accumulated using electronic reduction networks \cite{maeri2018}, to generate the final result. The \textit{psum} reduction latency and energy consumption are non-trivial components of overall latency and energy consumption \cite{STIFT}. Therefore, the value of \textit{N} plays a crucial role in governing the overall performance of DPUs. (2) The maximum achievable values of \textit{M} (fan-out degree/count of parallel DPEs). The value of \textit{M} directly decides the parallelism and consequently achieved throughput by a DPU. (3) The bit precision (\textit{B}) of input and weight values. If the supported value of \textit{B} is less than the precision requirement of GEMM operations, bit-slicing is applied to input and weight values \cite{HQNNA}. Due to bit-slicing, the overall count of dot product operations increases, degrading the throughput and energy efficiency \cite{sconna}. Therefore, the fundamental driver for achieving high performance from optical DPUslies in maximizing the values of \textit{N}, \textit{M}, and \textit{B}. %Typically, setting the value of \textit{N} equal to \textit{M} is the norm of the analog optical DPUs \cite{lukasscalability}.    

\begin{figure}[h] 
    \centering
    \includegraphics[scale=0.47]{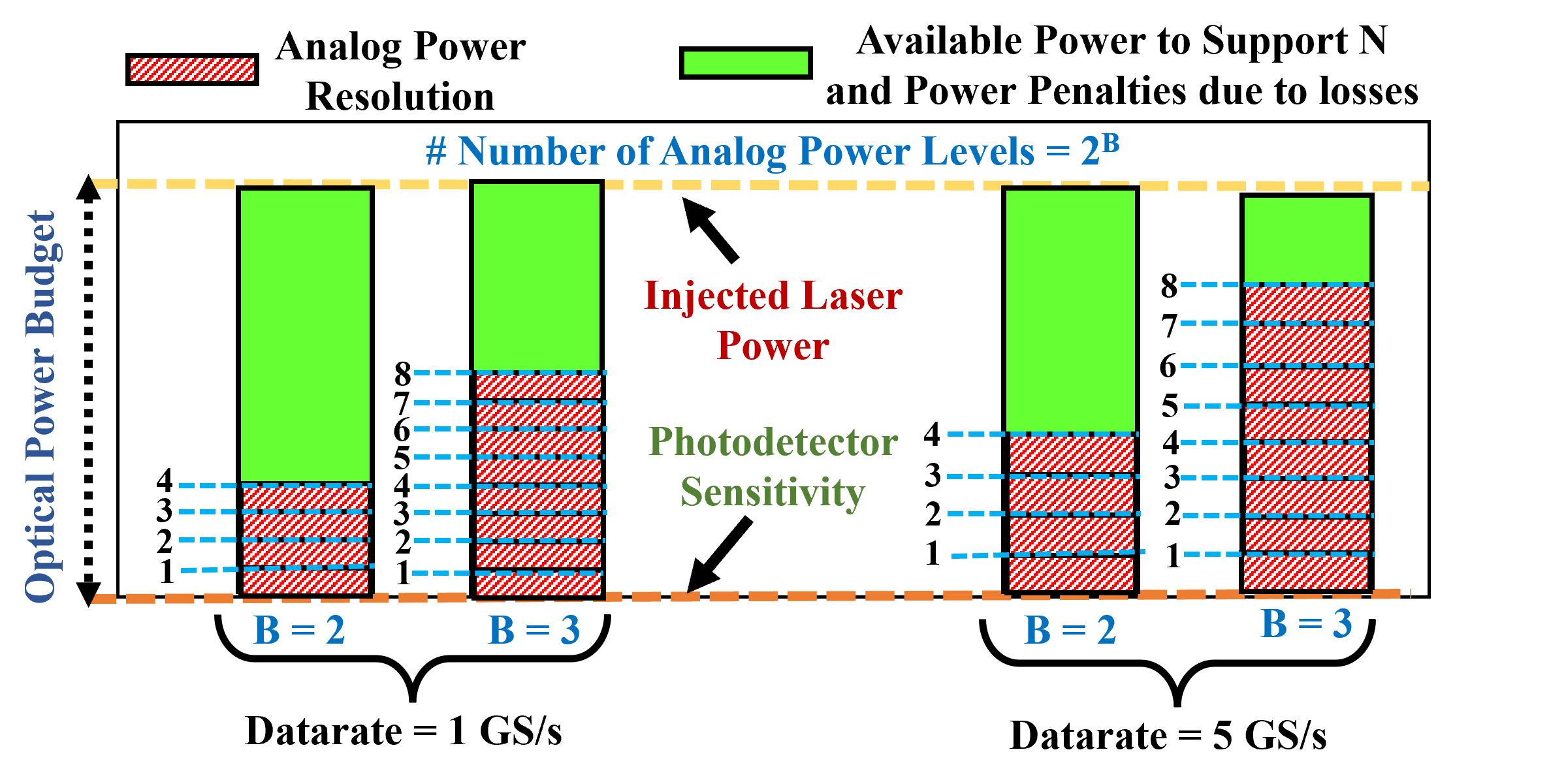}
    \caption{Conceptual breakdown of optical power budget usage and dependency of DPU size \textit{N} on supported bit precision \textit{B} for different values of \textit{B}=\{2, 3\}-bits across datarates DR=\{1, 5\} GS/s.}
    \label{fig3}
\end{figure}

In analog DPUs, a strong trade-off exists among supported values of \textit{N}, \textit{M}, and \textit{B} \cite{lukasscalability,cases2022}. The achievable values of \textit{M}, \textit{N}, and \textit{B} also strongly depend on the available optical power budget in the DPUs \cite{lukasscalability,cases2022}. This is illustrated in Fig. \ref{fig3}, assuming \textit{N}=\textit{M}, which is a common assumption in the literature \cite{lukasscalability,cases2022}. For the bit precision \textit{B}=2, 2$^B$=4 analog optical power levels are required that consume a large part of the available power budget, and the remaining power budget is used to support \textit{N} and power penalty (incurred due to crosstalk effects and signal losses) in the DPU. As \textit{B} increases to 3-bits, a larger part of the power budget is used to support \textit{B}, and the available power budget to support \textit{N} and power penalty further decreases. As a result, the supported value of \textit{N} decreases too. Similar impact can be observed when the operating datarate of DPUs increases (Fig. \ref{fig3}). Low \textit{N=M} decreases fan-in and fan-out degrees in the DPU, hampering the achievable throughput and energy efficiency. No prior work has characterized this impact, which has motivated this work. 

%the supported size of dot product operation and number of parallel dot product operations which in turn decreases the throughput. Furthermore, in analog optical DPUs, the increase in datarate diminishes the analog resolution, which further decreases the power available to support \textit{N} and losses in the DPU. Therefore, for a given \textit{B} and \textit{DR}, the achievable \textit{N} of the DPU depends on the total losses incurred by the DPU. The total losses in an optical DPU include optical signal losses and various crosstalk noises (more on the losses in Section \ref{sec4}). However, none of the prior works have analyzed the affect of DPU organization on various components of total losses and their impact on the achievable DPU size \textit{N}. In our work, we conduct an extensive circuit-level comparative analysis of ASMW, MASW, and SMWA organizations to assess how the organization of DPUs affects losses and, as a result, identify the achievable values of \textit{N}(=M) for various combinations of \textit{B} and \textit{DR}. We also evaluate the impact of the DPU organization on the overall performance in terms of throughput and energy efficiency.

\section{Circuit-Level Comparative Analysis}\label{sec4} 
In this section, we discuss the impact of DPU organization on power penalties due to various crosstalk effects and optical signal losses. 

% However, we determine that the presence or absence of various losses at these blocks depends on the organization of the DPUs. The \filledcircled{S} block losses consist of splitter insertion loss \cite{holylight} and propagation losses. The splitter insertion loss and propagation losses are present in the \filledcircled{S} block of all DPU organizations ASMW, MASW, and SMWA. In contrast to \filledcircled{S} block, the losses in the \filledcircled{A}, \filledcircled{M}, and \filledcircled{W} blocks are contingent on DPU organization. Next, we discuss the losses that are impacted by the DPU organization.   
              
\subsection{Impacts on Power Penalty Due to Crosstalk Effects}
\subsubsection{Inter-modulation crosstalk}
\begin{figure}[h] 
    \centering
    \includegraphics[scale=0.39]{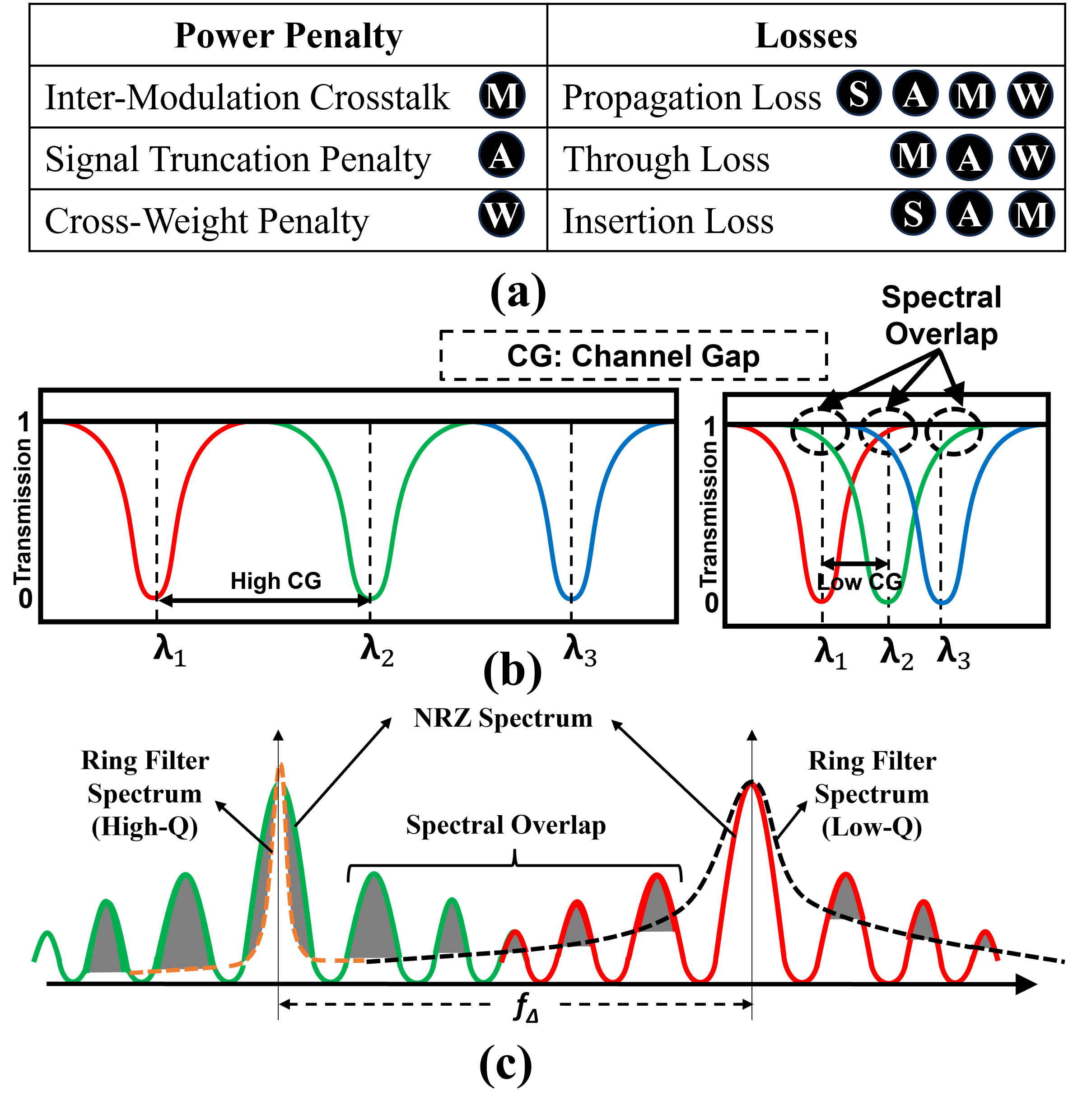}
    \caption{(a) Types of losses and power penalties at different optical signal manipulation blocks of optical DPUs. Illustration of (b) Inter-Modulation crosstalk at MRM input arrays \cite{padmaraju2014intermodulation,karempudi2022}, and (c) Filter crosstalk and signal truncation at filters \cite{filterpenalty}.}
    \label{fig4}
\end{figure}

From Fig. \ref{fig4}(a), inter-modulation crosstalk exists at the \filledcircled{M} block. Inter-modulation crosstalk occurs when an MRM unexpectedly modulates a neighboring wavelength channel instead of its assigned wavelength channel \cite{padmaraju2014intermodulation,karempudi2022}. Therefore, it is possible only if there are multiple wavelength channels present in the waveguide at a narrow channel spacing when the MRM is modulating its assigned wavelength channel (Fig. \ref{fig4}(b)).
%is present when MRMs at the input array are cascaded on a single waveguide containing multiple wavelength channels (refer to MRM input array in Fig. \ref{fig2}(b)). As shown in the transmission spectra on the left side of Fig. \ref{fig4}(b), the resonance wavelengths of the input MRMs are deliberately kept apart to avoid any overlap. However, as the number of wavelength channels increases and more input MRMs are cascaded, the channel gap between the resonances of the individual input MRMs decreases. As a result, the resonances will begin to overlap, as illustrated in Fig. \ref{fig4} (b) (on the right). Consequently, each input MRM not only modulates its corresponding wavelength channel but also modulates the neighboring wavelength channel. This phenomenon introduces inter-modulation crosstalk \cite{padmaraju2014intermodulation}\cite{karempudi2022} in cascaded input MRMs, as a result, the required power to compensate the crosstalk is defined as inter-modulation crosstalk power penalty. The inter-modulation crosstalk is present only when input MRMs operate on a waveguide with multiple wavelength channels. 
Hence, the necessary condition for inter-modulation crosstalk to occur is that the \filledcircled{M} block should appear after the \filledcircled{A} block in a DPU organization because, then only, the MRMs could have accessible neighboring wavelength channels to unexpectedly modulate them. Therefore, from Table \ref{table1}, inter-modulation crosstalk is present only in the ASMW DPUs.   

\subsubsection{Cross-weight penalty}
The MRR weight arrays in the \filledcircled{W} block can exhibit cross-weight penalty \cite{tait2016crossweightpenalty}, as shown in Fig. \ref{fig4}(a). Due to an insufficient channel gap between the adjacent optical wavelength channels, a weight MRR could perform undesired weighting on the neighboring optical wavelength channels leading to cross-weight penalty \cite{tait2016crossweightpenalty}. The additional power required to compensate for this crosstalk noise (to keep the signal-to-noise ratio intact) provides quantification of the cross-weight penalty \cite{tait2016crossweightpenalty}. Similar to inter-modulation crosstalk, the necessary condition for cross-weight penalty to occur is that the \filledcircled{W} block should appear after the \filledcircled{A} block in a DPU organization because, then only, the weight MRRs could have accessible neighboring wavelength channels to unexpectedly apply weighting to them. Therefore, from Table \ref{table1}, cross-weight penalty is present only in the ASMW and MASW DPUs.

%The cross-weight penalty at \filledcircled{W} block is present only when the weight MRRs operate on a single waveguide with multiple wavelength channels. Hence, a cross-weight penalty exists when the \filledcircled{W} block is arranged after the \filledcircled{A} block. Thus, it is present in ASMW DPUs and MASW DPUs as reported in Table \ref{table1}.  
 
\subsubsection{MRR Filter Penalty}
In general, the MRR filter penalty consists of two components \cite{filterpenalty}: (1) Optical losses due to signal truncation, and (2) inter-channel crosstalk. However, the inter-channel crosstalk component manifests only when an MRR filter is utilized in the demultiplexing configuration to demultiplex a signal from a waveguide containing multiple signals. Since demultiplexing is not required in our considered DPU organizations, the inter-channel crosstalk component remains absent from the MRR filter penalty discussed/analyzed in this paper. On the other hand, the optical losses due to signal truncation occur when a modulated optical signal is only partially transmitted through the MRR filter used as a multiplexer\cite{filterpenalty}. From Fig. \ref{fig4}(a), this phenomenon exists at the \filledcircled{A} block. Each N$\times$1 multiplexer consists of multiple MRR filters, and when the passband of a filter does not overlap perfectly with the passband of its corresponding modulated optical signal, it leads to the truncation of the signal sidelobes. Signal truncation is illustrated in Fig. \ref{fig4}(c) (on the left); when the filter has a high-quality factor (Q), its passband only partially overlaps with the passband of the modulated optical signal, resulting in signal truncation. Signal truncation occurs on modulated optical signals only; it does not occur on unmodulated optical wavelength channels. This is because unmodulated wavelength channels do that have any spectral sidelobes. Therefore, the necessary condition for filter penalty to occur is that the \filledcircled{A} block should appear after the \filledcircled{M} block in a DPU organization. The \filledcircled{A} block is organized after the \filledcircled{M} block in both MASW and SMWA DPUs, therefore signal truncation is present in these organizations (Table \ref{table1}).  
% This occurs due to the Lorentzian-shaped filtering response of the monowavelength filter, which has transitions that are not sharp. 

\begin{table}[]\centering
\caption{Crosstalks effects present in various DPU organizations.}
\label{table1}
\begin{tabular}{|c|c|c|c|}
\hline
                                    & \textbf{ASMW} & \textbf{MASW} & \textbf{SMWA} \\ \hline
\textbf{Inter-modulation Crosstalk} &      \checkmark         &       X        &        X       \\ \hline
\textbf{Cross-Weight Penalty}       &       \checkmark        &      \checkmark         &     X          \\ \hline
\textbf{Signal Truncation at Filters}       &       X        &      \checkmark          &     \checkmark           \\ \hline
\end{tabular}
\end{table}

\subsection{Impacts Due to Optical Signal Losses}
\subsubsection{Through losses}
The through losses are the optical power losses experienced by a wavelength channel as it traverses through MRMs and MRRs that are out-of-resonance to the wavelength channel in interest but operate on adjacent wavelength channels. From Fig. \ref{fig4}(a), through losses are present in blocks \filledcircled{M}, \filledcircled{W}, and \filledcircled{A}. The total amount of through loss experienced by a wavelength channel depends on the number of devices it traverses before reaching BPD. For instance in Fig. \ref{fig2}(b), after the splitter,  the wavelength $\lambda_{1}$(indicated with the color red) passes through \textit{(N-1)} out-of-resonance MRMs and \textit{(N-1)} out-of-resonance MRRs before reaching the BPD. The reduction in optical power of $\lambda_{1}$ as it interacts with these devices is termed as its through loss. The total through losses can be determined by summing the individual through losses caused by individual MRRs and MRMs. The through losses vary across DPU organizations as reported in Table \ref{table2}. For example, from Fig. \ref{fig2}(b), Fig. \ref{fig2}(c), and Fig. \ref{fig2}(d) the total number out-of-resonance MRMs and MRRs traversed by $\lambda_1$ are \textit{2(N-1)}, \textit{N}, and 2 in ASMW, MASW, and SMWA DPUs respectively. Therefore,  $\lambda_1$ incurs higher through losses in ASMW DPUs than MASW and SMWA DPU organizations. Similarly, other optical wavelength channels $\lambda_2-\lambda_N$ also experience higher through losses in ASMW DPUs.

\subsubsection{Insertion losses}
The insertion losses are the optical power losses encountered by a wavelength channel while devices such as MRMs, MRRs, and filters operate on it. The total insertion losses experienced by wavelength channels are approximately the same across DPU organizations.  

% Consequently, the through loss in a DPU is dependent on the total number of devices that an optical wavelength interacts with before reaching the BPD. The total number of devices along the path of an optical wavelength channel is determined by the specific organization of the DPU.

% From Fig. \ref{fig2}(b), Fig. \ref{fig2}(c), and Fig. \ref{fig2}(d), the total number of devices (including MRMs, MRRs, and monowavelength filters) traversed by $\lambda_1$ are \textit{2N}, \textit{N}, and \textit{N+2} in ASMW, MASW, and SMWA, respectively. Therefore,  $\lambda_1$ incurs higher through losses in ASMW compared to MASW and SMWA organizations. Similarly, other optical wavelength channels $\lambda_2-\lambda_N$ also experience higher through losses in ASMW organization. From Fig. \ref{fig2}(d), the traversal path of $\lambda_1$ in SMWA encounters a total \textit{N+2} devices consisting of 1 MRM at MRM weight array, 1 MRR at MRR input array, and \textit{N} monowavelength filters at aggregation lanes, which is higher compared to MSWA (refer Fig. \ref{fig2}(c)) with a total of \textit{N+1} devices consisting of 1 MRM at MRM input array and \textit{N} MRRs at the MRR weight array. However, the through loss incurred by a monowavelength filter is less than that of MRR, therefore MASW with more MRRs experiences high overall through losses compared to SMWA. 

\subsubsection{Waveguide Propagation Losses}
Propagation losses are the sum of scattering losses (due to the sidewall roughness of the
waveguide) and absorption losses (due to the material and free-carrier absorption mechanisms in the waveguide).  From Fig.\ref{fig4}(a), propagation losses are present in all the blocks. Propagation losses increase proportionally with the length of the waveguide. The SMWA DPUs, because of their spectrally hitless architecture, employ a larger number of longer waveguides \cite{karen2020proceeding}, resulting in increased propagation losses compared to ASMW and MASW DPUs. Additionally, MASW DPUs experience lower propagation losses than ASMW DPUs. This is because the MRR weight arrays in MASW DPUs share a single MRM input array, which reduces the overall waveguide length and corresponding propagation losses.

%Tables \ref{table1} and \ref{table2} summarize our comparative analysis and discussion. The total losses in a DPU vary based on the organization and lead to variation in the achievable DPU size \textit{N(=M)}.
\begin{table}[]
\caption{Optical losses present in various DPU organizations.}
\label{table2}\centering
\begin{tabular}{|c|c|c|c|}
\hline
\multicolumn{1}{|l|}{}                & \textbf{ASMW}                             & \textbf{MASW}                             & \textbf{SMWA}                             \\ \hline
\textbf{Through Losses}               & \cellcolor[HTML]{FD6864}\textbf{High}     & \cellcolor[HTML]{FFCC67}\textbf{Moderate} & \cellcolor[HTML]{67FD9A}\textbf{Low}      \\ \hline
\textbf{Propagation Losses}           & \cellcolor[HTML]{FFCC67}\textbf{Moderate} & \cellcolor[HTML]{67FD9A}\textbf{Low}      & \cellcolor[HTML]{FD6864}\textbf{High}     \\ \hline
\end{tabular}
\end{table}

\begin{table}[]
\centering
\caption{Definition and values of various parameters used in Eq. \ref{eq3}, Eq. \ref{eq4}, and Eq. \ref{eq5} (from \cite{lukasscalability}) for the scalability analysis. }
\label{abbrevations}
\begin{tabular}{|c|c|c|}
\hline
{ \textbf{Parameter}}             & { \textbf{Definition}}                                                                                                         & { \textbf{Value}}   \\ \hline
{$P_{Laser}$}                & { Laser Power Intensity}                                                                                               & { 10 dBm}  \\ \hline
 { $R_{s}$}                     & { PD Responsivity}                                                                                                     & { 1.2 A/W} \\ \hline
{ $R_L$}                    & { Load Resistance}                                                                                                     & { 50 $\Omega$}      \\ \hline
{ $I_d$}                    & { Dark Current}                                                                                                        & { 35 nA}   \\ \hline
 { T}                     & { Absolute Temperature}                                                                                                & { 300 K}   
\\ \hline
{ RIN}                   & { Relative Intensity Noise}                                                                                            & { -140 dB/Hz}   \\ \hline 
{ $P_{EC-IL}$}            &\begin{tabular}[c]{@{}c@{}}Fiber to Chip Coupling \\ Insertion Loss\end{tabular}                                                                                & { 1.44 }     \\ \hline
{ $P_{MRR-W-IL}$}         & \begin{tabular}[c]{@{}c@{}} Silicon Waveguide  \\ Insertion Loss\end{tabular}                                                                                   & { 0.3 dB/mm}     \\ \hline
 { $P_{splitter-IL}$}      & { Splitter Insertion Loss}                                                                                             & { 0.01 dB}    \\ \hline
 { $P_{MRM-IL}$}           & \begin{tabular}[c]{@{}c@{}}Optical Microring Modulator  \\ Insertion Loss\end{tabular}                                                                           & { 4 dB}       \\ \hline
  { $P_{MRR-IL}$}           & \begin{tabular}[c]{@{}c@{}}Optical Microring Resonator  \\ Insertion Loss\end{tabular}                                                                           & { 0.01 dB}       \\ \hline
 { $P_{MRM-OBL}$}          & { Out of Band Loss}                                                                                                & { 0.01 dB}                                                                                 \\ \hline
 \multirow{0}{*}{$P_{Penalty}$} & MASW Network Penalty  & 4.8 dB\\ \cline{2-3} 
                    & ASMW Network Penalty  & 5.8 dB \\ \cline{2-3} 
                    & SMWA Network Penalty & 1.8 dB\\ \hline

\end{tabular}
\end{table}

\subsection{Scalability Analysis}\label{scalabilityanalysis}
\begin{figure}[h]
  \centering
  \includegraphics[width=\linewidth]{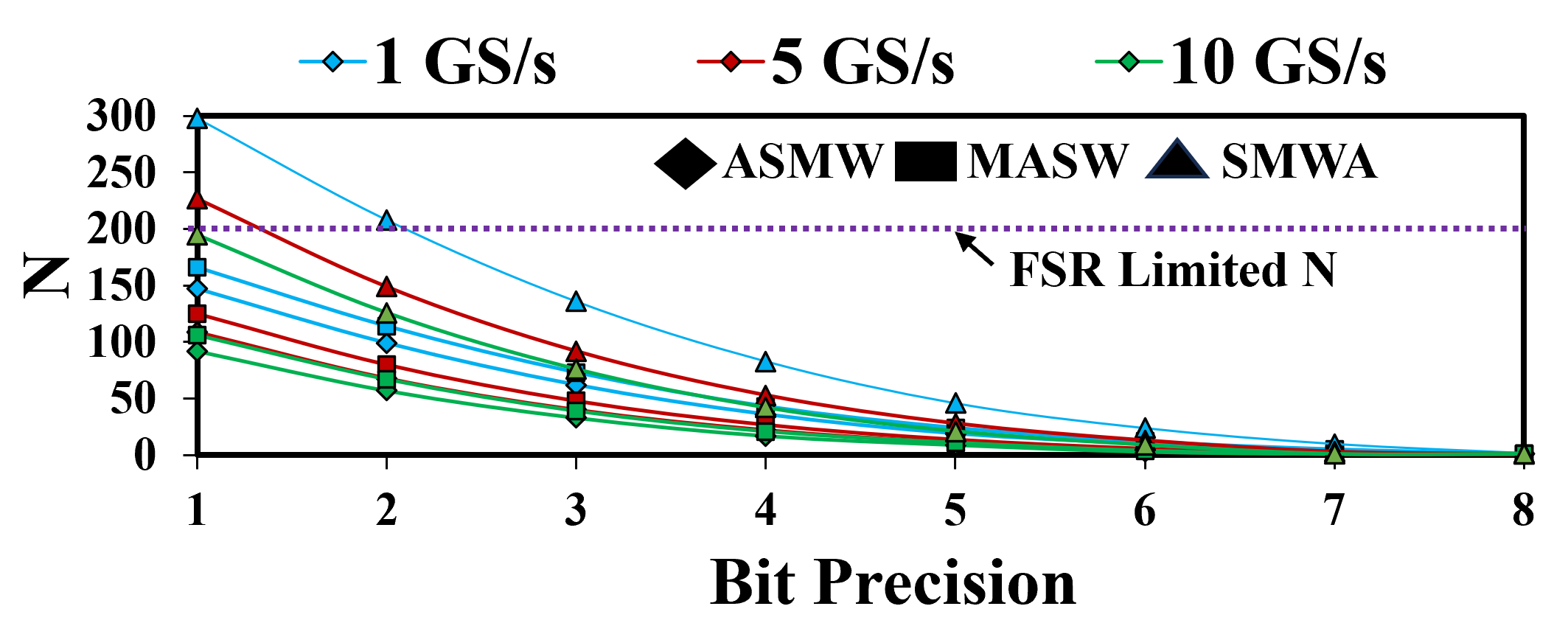}
  \caption{Supported DPU size N (=M) for bit precision=\{1, 2, 3, 4, 5, 6, 7, 8\} bits at data rates (DRs)=\{1, 5, 10\} GS/s, for AMW, MAW, and MWA DPUs.} 
  \label{fig6}
\end{figure}

To determine the achievable size \textit{N} for our ASMW, MASW, and SMWA DPU organizations, we adopt scalability analysis equations (Eq. \ref{eq3}, Eq. \ref{eq4}, and Eq. \ref{eq5}) from \cite{lukasscalability} and \cite{cases2022}. Table \ref{abbrevations} reports the definitions of the parameters and their values used in these equations. The reported $P_{Penalty}$ includes the summation of inter-modulation crosstalk, cross-weight penalty, filter penalty, and propagation losses. We consider optimistic values for these parameters, with inter-modulation crosstalk of  $\leq$1dB, cross-weight penalty of $\leq$3dB, and filter penalty of $\leq$0.5dB. To achieve such optimistic inter-modulation crosstalk and cross-weight penalties the channel spacing should be equal to 0.4$\times$FWHM \cite{tait2016crossweightpenalty}. We considered Free Spectral Range (FSR=50nm) \cite{lukasscalability} with FWHM=0.7nm, resulting in channel spacing of 0.25nm(=0.4$\times$0.7). Then, the FSR limited N value is 200(=FSR/0.25). 
  We consider \textit{M=N} and first solve Eq. \ref{eq3} and Eq. \ref{eq4} for a set of DRs=\{1, 5, 10\} GS/s, to find a corresponding set of $P_{PD-opt}$. Then, we solve Eq. \ref{eq5} for the maximum value of \textit{N} that achieves $P_{O/p}$  greater than obtained $P_{PD-opt}$ values across the set of \textit{DR}s. Fig. \ref{fig6} reports the achievable \textit{N} of ASMW, MASW, and SMWA DPUs for different bit-precision levels (B) across various \textit{DR}s.  The achievable \textit{N} value defines the feasible number of MRRs per DPE; thus, this \textit{N} also defines the maximum size of the dot product that can be generated in our DPU. As evident from Fig.\ref{fig6}, SMWA can support larger \textit{N} value compared to ASMW and MASW at all bit-precision levels across different DRs. For instance, SMWA achieves larger \textit{N=83} for 4-bit precision at 1 GS/s, compared to ASMW and MASW, which achieve \textit{N=36} and \textit{N=43}, respectively. This is because of SMWA's DPU  architecture, as reported in Table \ref{table1}, SMWA significantly reduces crosstalk-related power penalty reducing $P_{penalty}$. This enables SMWA to support larger \textit{N} compared to ASMW and MASW at the same input laser power. 

\begin{equation}
       B = \frac{1}{6.02}\Bigg[20log_{10}(\frac{R\times P_{PD-opt}}{\beta\sqrt{\frac{DR}{\sqrt{2}}}}-1.76\Bigg]
       \label{eq3}
\end{equation}

\begin{dmath}
    \beta = \sqrt{2q(RP_{PD-opt}+I_d)+\frac{4kT}{R_L}+R^2P_{PD-opt}^2RIN} +\sqrt{2qI_d+\frac{4kT}{R_L}}
    \label{eq4}
\end{dmath}
    
\begin{dmath}
\label{eq5}
     P_{O/p}(dBm) = P_{Laser}-P_{SMF-att}-P_{EC-IL}-P_{Si-att}\times N\times d_{MRR}-P_{MRM-IL}-(N-1)P_{MRM-OBL}-P_{splitter-IL}\times log_{2}(M)-P_{MRR-W-IL}-(N-1)P_{MRR-W-OBL}-P_{penalty}-10log_{10}(N)
\end{dmath}

\section{Evaluation}\label{sec5}
\subsection{System Level Implementation}
\begin{figure}[h]
  \centering
  \includegraphics[scale=0.3]{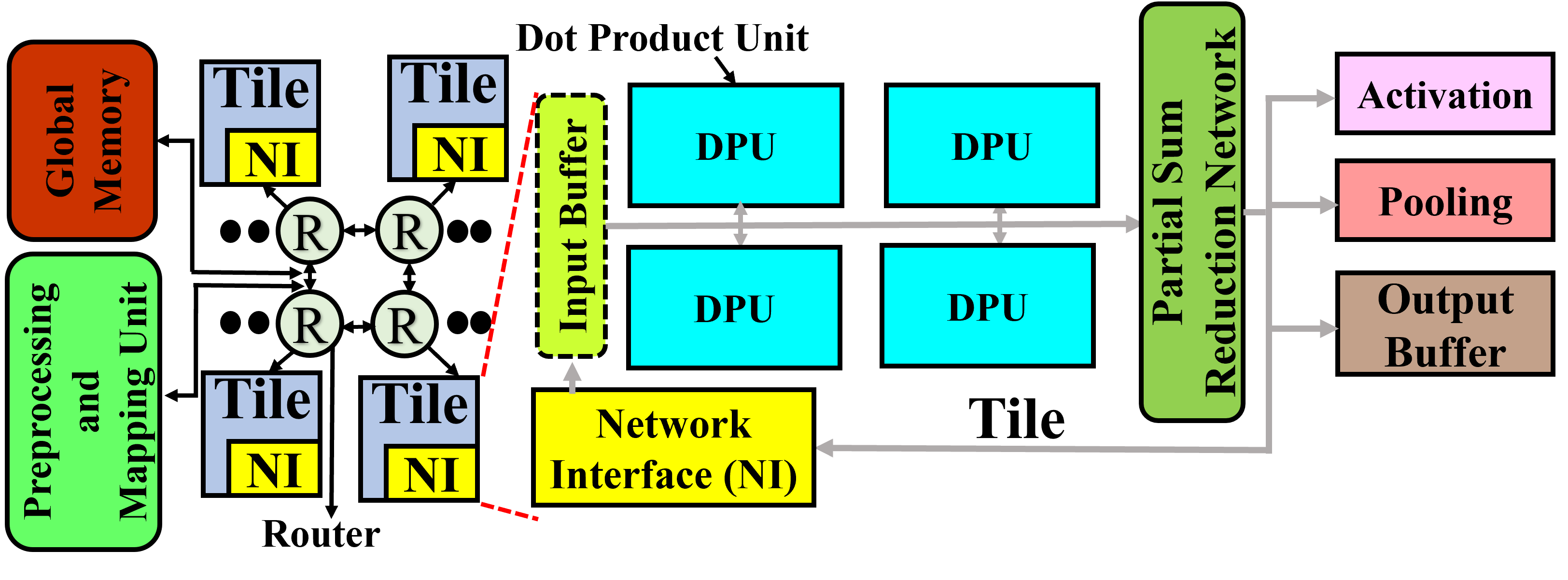}
  \caption{System-level overview of Photonic GEMM accelerator.} 
  \label{systemlevelimplement}

\end{figure}

Fig. \ref{systemlevelimplement} illustrates the general system-level implementation of incoherent photonic GEMM accelerators. It consists of global memory that stores CNN parameters and a pre-processing and mapping unit. It has a mesh network of tiles. Each tile contains 4 DPUs interconnected (via H-tree) with a unified buffer as well as pooling and activation units. Each DPU consists of multiple DPEs and each DPE is equipped with a dedicated input and output FIFO buffer \cite{Wang2021-ISCAS} to store intermittent weights, inputs, and \textit{psums} values. In addition, each tile also contains a \textit{psum} reduction network.

\begin{table}[]
\caption{DPU size (N) and DPU Count (\#) at 4-bit precision across various DRs for
different accelerators architectures.}
\label{scalabilityandareaproportionate}\centering
\begin{tabular}{|c|cccccc|}
\hline
\multicolumn{1}{|l|}{} & \multicolumn{6}{c|}{\textbf{Datarate}}                                                                                                   \\ \hline
                       & \multicolumn{2}{c|}{\textbf{1 GS/s}}                        & \multicolumn{2}{c|}{\textbf{5 GS/s}}                        & \multicolumn{2}{c|}{\textbf{10 GS/s}}   \\ \hline
\textbf{DPU} &
  \multicolumn{1}{c|}{\textbf{N}} &
  \multicolumn{1}{c|}{\textbf{\#}} &
  \multicolumn{1}{c|}{\textbf{N}} &
  \multicolumn{1}{c|}{\textbf{\#}} &
  \multicolumn{1}{c|}{\textbf{N}} &
  \textbf{\#} \\ \hline
\textbf{ASMW}           & \multicolumn{1}{c|}{36} & \multicolumn{1}{c|}{160} & \multicolumn{1}{c|}{17} & \multicolumn{1}{c|}{265} & \multicolumn{1}{c|}{12} & 291 \\ \hline
\textbf{MASW}           & \multicolumn{1}{c|}{43} & \multicolumn{1}{c|}{186} & \multicolumn{1}{c|}{21} & \multicolumn{1}{c|}{275} & \multicolumn{1}{c|}{15} & 295 \\ \hline
\textbf{SMWA}         & \multicolumn{1}{c|}{83} & \multicolumn{1}{c|}{50}  & \multicolumn{1}{c|}{42} & \multicolumn{1}{c|}{147} & \multicolumn{1}{c|}{30} & 198  \\ \hline
\end{tabular}
\end{table}

\begin{table}[]
\begin{threeparttable}[b]
\caption{Accelerator Peripherals and DPU Parameters {\cite{cases2022}}}
\label{acceleratorparameters}
\begin{tabular}{|c|ccc|}
\hline
                           & \multicolumn{1}{c|}{\textbf{Power(mW)}} & \multicolumn{1}{c|}{\textbf{Latency}} & \textbf{Area($mm^2$)} \\ \hline
\textbf{Reduction Network} & \multicolumn{1}{c|}{0.050}              & \multicolumn{1}{c|}{3.125ns}          & 3.00E-5            \\ \hline
\textbf{Activation Unit}   & \multicolumn{1}{c|}{0.52}               & \multicolumn{1}{c|}{0.78ns}           & 6.00E-5            \\ \hline
\textbf{IO Interface}      & \multicolumn{1}{c|}{140.18}             & \multicolumn{1}{c|}{0.78ns}           & 2.44E-2            \\ \hline
\textbf{Pooling Unit}      & \multicolumn{1}{c|}{0.4}                & \multicolumn{1}{c|}{3.125ns}          & 2.40E-4            \\ \hline
\textbf{eDRAM}             & \multicolumn{1}{c|}{41.1}               & \multicolumn{1}{c|}{1.56ns}           & 1.66E-1             \\ \hline
\textbf{Bus}               & \multicolumn{1}{c|}{7}                  & \multicolumn{1}{c|}{5 cycles}         & 9.00E-3               \\ \hline
\textbf{Router}            & \multicolumn{1}{c|}{42}                 & \multicolumn{1}{c|}{2 cycles}         & 1.50E-2              \\ \hline
\textbf{DAC \cite{dac1} }      & \multicolumn{1}{c|}{12.5}         & \multicolumn{1}{c|}{0.78ns}             & 2.50E-3              \\ \hline
\textbf{ADC(1 GS/s) \cite{adc1gbps} }      & \multicolumn{1}{c|}{2.55}         & \multicolumn{1}{c|}{0.78ns}             & 2E-3              \\ \hline
\textbf{ADC(5 GS/s) \cite{adc3gbps} }      & \multicolumn{1}{c|}{11}         & \multicolumn{1}{c|}{0.78ns}             & 21E-3              \\ \hline
\textbf{ADC(10 GS/s) \cite{adc5gbps} }      & \multicolumn{1}{c|}{30}         & \multicolumn{1}{c|}{0.78ns}             & 103E-3           \\ \hline
\textbf{EO Tuning}         & \multicolumn{1}{c|}{80 $\mu$W/FSR}          & \multicolumn{1}{c|}{20ns}             & -                  \\\hline
\textbf{TO Tuning}         & \multicolumn{1}{c|}{275 mW/FSR}         & \multicolumn{1}{c|}{4$\mu$s}             & -                  \\ \hline

\end{tabular}
% \begin{tablenotes}
%        \item $^1$All accelerators operate DAC at 1 GS/s.
%        \item $^2$HEANA-OS operate DAC at 10 GS/s.

% \end{tablenotes}
  \end{threeparttable}
\end{table}

\subsection{Simulation Setup}
For evaluation, we model system-level implementation of AMSW, MASW, and SMWA GEMM accelerator architectures using our developed custom, transaction-level, event-driven Python-based simulator. Using the simulator, we simulated the inference of four CNN models (with a batch size of 1): GoogleNet\cite{googlenet}, ResNet50\cite{resnet}, MobileNet\_V2 \cite{mobilenetv2}, and ShuffleNet\_V2 \cite{shufflenet}. We evaluate the metrics such as Frames per second (FPS), FPS/W (energy efficiency), and  FPS/W/mm$^2$ (area efficiency). Further, prior optical GEMM accelerators \cite{deapcnn,holylight, crosslight} show minimal or no loss in inference accuracy.

We compared AMSW, MASW, and SMWA accelerator architectures for inference of 8-bit integer quantized CNN models. All accelerators are operated for 4-bit integer precision across data rates 1GS/s, 5GS/s, and 10GS/s, from Fig. \ref{fig6}, for these parameters SMWA, ASMW, and MASW achieve \textit{N} reported in Table \ref{scalabilityandareaproportionate}, respectively. We omitted
comparison with CMOS-based digital CNN accelerators as prior photonic accelerators (e.g., \cite{deapcnn, holylight, crosslight}) have already outperformed them. 
 Our evaluation is based on output stationary dataflow. For a fair comparison, we performed area proportionate analysis, wherein we altered the DPU count for each photonic CNN accelerator across all of the accelerator's DPUs to match with the area of SMWA (\textit{N=83}) having 50 DPUs. Table \ref{scalabilityandareaproportionate} reports the scaled DPU count of ASMW, MASW, and SMWA across various datarates.  

Table \ref{acceleratorparameters} gives the parameters used for evaluating the overhead of the peripherals in our evaluated accelerators. We consider each laser diode to emit input optical power of 10 mW (10 dBm) (Table \ref{abbrevations})\cite{deapcnn}, multiplexer and splitter parameters are taken from \cite{holylight}. 

\subsection{Evaluation Results}
\begin{figure}[h]
  \centering
  \includegraphics[width=\linewidth]{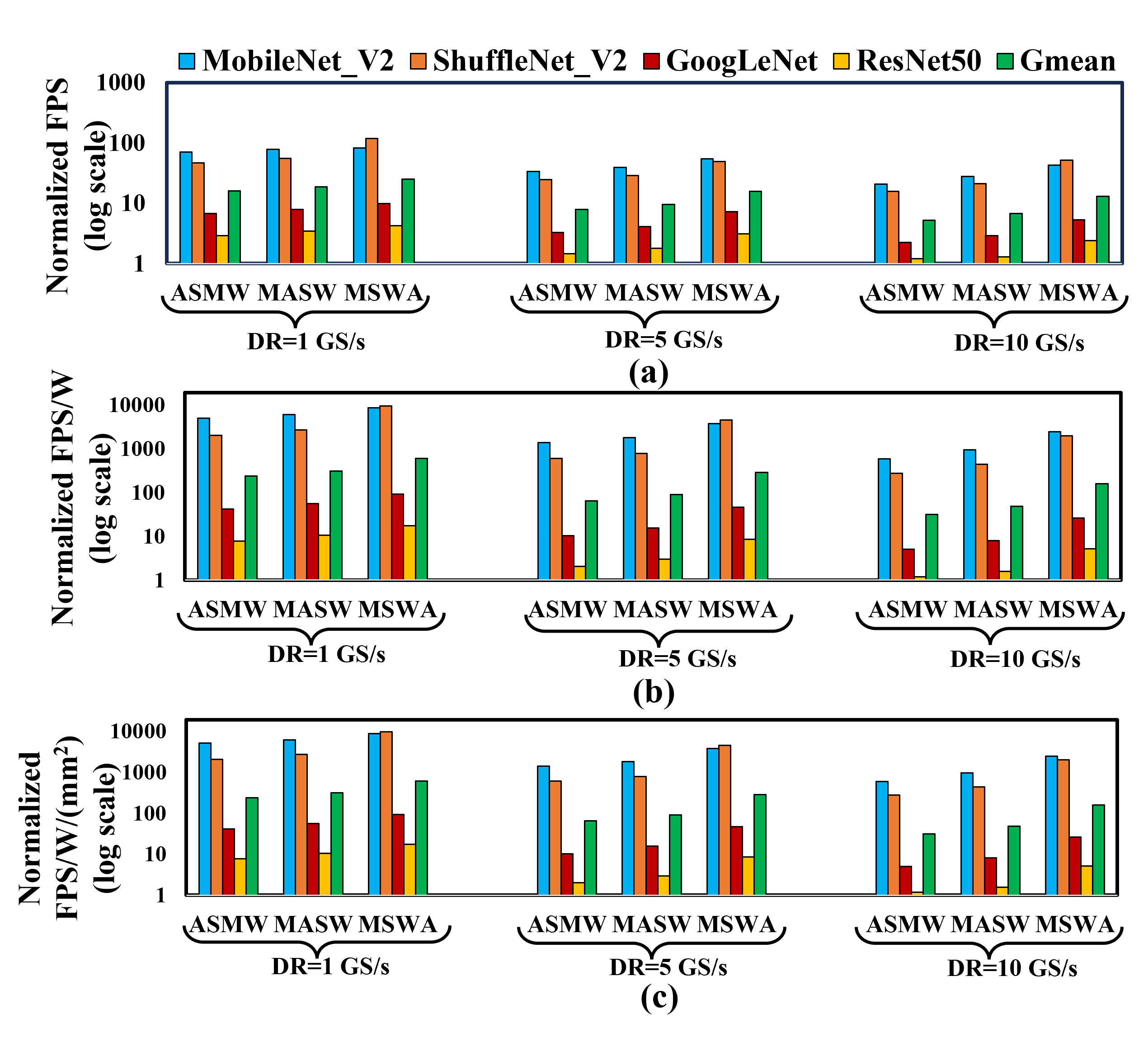}
  \caption{(a) Normalized FPS (log scale) (b) Normalized FPS/W (log scale) (c) Normalized FPS/W/mm$^2$ (log scale)\textbf{} for AMW, MAW, and MWA accelerators with input batch size=1. Results of FPS, FPS/W, FPS/W/mm$^2$ are normalized with respect to AMW executing ResNet50 at 10 GS/s.} 
  \label{fpsandenergyefficieny}
\end{figure}

Fig. \ref{fpsandenergyefficieny}(a) shows Normalized FPS results for various accelerators with batch size=1 at different datarates (DRs), normalized to ASMW for ResNet50 at 10 GS/s. SMWA accelerator outperforms MASW and ASMW accelerators, respectively, on gmean across four CNN models for all data rates. At 1 GS/s, SMWA achieves up to 2.5$\times$ and 2.3$\times$ better FPS than ASMW and MASW, respectively. As DR increases to 5 GS/s and 10 GS/s, SMWA shows better improvements in FPS, achieving up to 3.9$\times$ and 4.4$\times$ better FPS than ASMW, respectively. Similarly, SMWA achieves up to 3.6$\times$ and 3.9$\times$ better FPS than MASW at 5 GS/s and 10 GS/s. 

These significant improvements in throughput for SMWA are due to two reasons. First, inter-modulation crosstalk and cross-weight related power penalties are absent in SMWA (refer Table. \ref{table1}) due to its hitless architecture. This allows SMWA to support a larger DPU size (\textit{N=83}), i.e., the size of the dot product operation \textit{N} (refer Table \ref{scalabilityandareaproportionate}) and the number of parallel dot product operations \textit{M} (=\textit{N}). Consequently, the overall throughput is increased with improved parallelism.  Secondly, larger \textit{N} generates less number of \textit{psums} which reduces the use of a partial sum reduction network. This, in turn, reduces the latency associated with \textit{psum} reductions and improves FPS. Among ASMW and MASW, MASW performs slightly better than ASMW at all datarates. MASW with input array sharing mitigates inter-modulation crosstalk power penalty at MRM input array and also incurs lower through losses compared to ASMW (refer Table \ref{table2}), due to these benefits MASW achieves slightly better \textit{N} compared to ASMW (refer Table \ref{scalabilityandareaproportionate}). MASW with higher \textit{N} achieves better parallelism and decreases reduction latency, resulting in better FPS.            

Furthermore, as datarate increases the FPS of each accelerator decreases, at 5GS/s and 10 GS/s, the value of \textit{N} decreases for all the accelerators (refer Table \ref{scalabilityandareaproportionate}) which results in a higher number of \textit{psums}. Therefore, the latency corresponding to \textit{psum} reduction increases with an increase in datarate and leads to lower FPS for accelerators. Overall, SMWA architecture with higher \textit{N} achieves better throughput than ASMW and MASW architectures.

Fig. \ref{fpsandenergyefficieny}(b) shows FPS/W (log scale) results for ASMW, MASW, and SMWA accelerator with batch size=1 at different DRs, normalized to ASMW for ResNet50 at 10 GS/s.
It is evident that the SMWA accelerator attains better energy efficiency than the MASW and ASMW accelerators. At 1 GS/s, SMWA gains 1.9$\times$ and 2.5$\times$ better FPS/W against analog MASM and ASMW, respectively, on gmean across the CNNs. As the datarate increases to 5 GS/s and 10 GS/s, SWMA achieves 3.17$\times$ and 3.3$\times$ improvements in FPS/W when compared to MASW. SMWA also exhibits a significant 4.4$\times$ and 5$\times$ improvement in FPS/W when compared to ASMW at 5 GS/s and 10 GS/s. These energy efficiency benefits of SMWA are due to the improved throughput and decreased energy consumption of \textit{psum} reductions. As discussed earlier, superior \textit{N} supported by SMWA improves parallelism which decreases dynamic energy consumption with improved throughput. In addition, SMWA also requires the least number of \textit{psum} reductions and this provides energy savings by reducing the usage of \textit{psum} reduction network. At higher datarates of 5 GS/s and 10 GS/s, the \textit{N} value decreases, consequently requiring more \textit{psum} reductions and \textit{psum} reduction energy consumption.  Furthermore, as datarate increases the accelerator peripherals like ADCs consume more static power (refer Table \ref{acceleratorparameters}) which also decreases the FPS/W achieved by each accelerator. Thus, as datarate increase the FPS/W decreases for ASMW, MASW, and SMWA accelerators. Overall,  SMWA provides better energy efficiency compared to the other accelerators.               

Fig. \ref{fpsandenergyefficieny}(c) shows the area efficiency values (FPS/W/mm$^2$) for each accelerator across various CNNs. The area efficiency results look similar to energy efficiency as we match the area of all the accelerators to SMWA (for the area proportionate analysis). SMWA gains up to 5.2 $\times$ and 3.4$\times$ better FPS/W/mm$^2$ against ASMW and MSAW, respectively, on gmean across four CNN models for all data rates. Overall, the SWMA accelerator achieves the best throughput, energy efficiency, and area efficiency. 

\section{Summary}
In this paper, we introduced a systematic approach for classifying prior incoherent MRR-based GEMM accelerators into three distinct categories based on their organization of optical signal manipulation blocks: (1) Modulation-Aggregation-Splitting-Weighting (MASW), (2) Aggregation-Splitting-Modulation-Weighting (ASMW), and (3) Splitting-Modulation-Weighting-Aggregation (SMWA). We performed a comprehensive circuit-level comparative analysis of MASW, ASMW, and SMWA organizations and identified that each organization incurs different magnitudes of crosstalk noise and optical signal losses. As a result, our scalability analysis at the circuit level demonstrated that each organization achieves different levels of processing parallelism. At the system level, our evaluation results for four CNN models show that SMWA organization achieves up to 4.4$\times$, 5$\times$, and 5.2$\times$ better throughput, energy efficiency, and area-energy efficiency, respectively, compared to ASMW and MASW organizations on average.

\section*{Acknowledgments}
We thank the anonymous reviewers whose valuable feedback helped us improve this paper. We would also like to acknowledge the National Science Foundation (NSF) as this research was supported by NSF under grant CNS-2139167.

\bibliographystyle{IEEEtran}
\bibliography{references}

\end{document}